\def\be{\begin{equation}}
\def\ee{\end{equation}}
\def\ba{\begin{array}}
\def\ea{\end{array}}

\documentclass[twocolumn,pra,showpacs,amsmath]{revtex4-1}
\def\qed{\leavevmode\unskip\penalty9999 \hbox{}\nobreak\hfill
     \quad\hbox{\leavevmode  \hbox to.77778em{%
               \hfil\vrule   \vbox to.675em%
               {\hrule width.6em\vfil\hrule}\vrule\hfil}}
     \par\vskip3pt}
\usepackage{amsmath}
\usepackage{epsfig}
\newtheorem{theorem}{Theorem}

\begin{document}

\title{Identification of three-qubit entanglement}

\author{Ming-Jing Zhao$^{1}$}
\author{Ting-Gui Zhang$^{1}$}
\author{Xianqing Li-Jost$^{1}$}
\author{Shao-Ming Fei$^{1,2}$}
\affiliation{$^1$Max-Planck-Institute for Mathematics in the Sciences,
Leipzig 04103, Germany\\
$^2$School of Mathematical Sciences, Capital Normal
University, Beijing 100048, China}

\begin{abstract}
We present a way in identifying all kinds of entanglement for
three-qubit pure states in terms of the expectation values of Pauli operators. The necessary and sufficient conditions to classify the
fully separable, biseparable and genuine entangled states are
explicitly given. The approach can be generalized to multipartite
high dimensional case. For three-qubit mixed states, we propose two kinds of inequalities in terms of the expectation values of complementary observables. One inequality has advantages in entanglement detection of quantum state with positive partial transpositions and the other is able to detect genuine entanglement. The results give an effective way in
experimental entanglement identification.
\end{abstract}

\maketitle

\section{Introduction}

Entanglement is the essential resource for many
tasks in quantum information processing \cite{M.A.Nielsen,R.
Horodecki}. As a result, various approaches have been proposed to
characterize entanglement. But there are yet no
operational necessary and sufficient separability criteria for
high dimensional states and multipartite states generally.

For unknown quantum states, the separability can only be determined
by measuring some suitable quantum mechanical observables. An
important approach to characterize entanglement is Bell inequality
\cite{Bell,Gisin,S. K. Choudhary,J. L. Chen,etc, S. Yu2012}. For instance, N.
Gisin has proved that all two-qubit pure entangled states violate
the CHSH inequality \cite{Gisin}. In three-qubit system, Ref.
\cite{J. L. Chen} presents a Bell-type inequality that would be
violated by all three-qubit pure entangled states
and Ref. \cite{S. K. Choudhary} shows analytically that all pure
entangled states violate another Bell-type inequality by exploiting the Hardy's nonlocality argument. Quite recently, Ref. \cite{S. Yu2012} shows all multipartite high dimensional entangled pure states violate a single Bell inequality. For general
mixed two-qubit states, Bell-type inequality has been proposed to
give the necessary and sufficient criterion of separability \cite{S.
Yu, mj-2byd inequ}. Besides Bell inequality, the entanglement
witness could also be used for experimental detection of quantum
entanglement for some special states, such as W state \cite{O. Guhne2003}, GHZ state \cite{O. Guhne2003} and cluster state \cite{Y. Tokunaga}. Some of these witnesses can be
implemented with the present technology \cite{O. Guhne2009, M. Bourennane}.
Another method to detect entanglement is to measure the entanglement
measures experimentally \cite{F. Mintert2005,S. P. Walborn2006,S. P.
Walborn2007,zmj-concurrence}, which have been implemented for
two-qubit pure state \cite{S. P. Walborn2006,S. P. Walborn2007}.

In multipartite systems, there are many kinds of entanglement. For the simplest case, in three-qubit system, all pure states are classified into six types in terms of stochastic local operations and classical communication (SLOCC) equivalence \cite{W. Dur}. They can also be classified into nine types by canonical form of pure three-qubit state \cite{A. Acin}. Then for three-qubit mixed states, they could be classified into four types if one demands each type consists a compact and convex set \cite{A. Acin2001}.
Different types of entanglement have different features. But it is general difficult to characterize different types
of multipartite entanglement and distinguish them from each other completely.
Entanglement witness and Bell inequality have been proposed to
distinguish important classes of qubit states \cite{C. Schmid, M.
Huber,B. Jungnitsch, N. Brunner}.

In this paper we mainly deal with the separability of quantum states
and distinguish different entanglement in three-qubit system. We
first express the bipartite entanglement of three-qubit pure state
in terms of expectation values of Pauli operators. Based on this,
we derive some inequalities which can be viewed as entanglement
witness to detect the separability of three-qubit pure states
completely. Therefore one can recognize whether a three-qubit pure
state is fully separable, biseparable or genuine entangled by
measuring some particular expectation values. For the entanglement detection of
three-qubit mixed states, we propose two kinds of inequalities in terms of the expectation values of complementary observables. One inequality is able to detect entanglement in quantum state with positive partial transpositions (PPT) and the other is able to detect genuine entanglement. These inequalities may help entanglement detection and differentiation in three-qubit system experimentally.

The paper is organized as follows. In section II, we express the bipartite entanglement of three-qubit pure state
in terms of expectation values of Pauli operators. Then necessary and sufficient conditions to classify the
fully separable, biseparable and genuine entangled states are
explicitly given for three-qubit pure state. In section III, we provide two kinds of inequalities in terms of the expectation values of complementary observables to detect entanglement in three-qubit mixed state. These inequalities are showed to have the ability to detect some PPT entanglement and genuine entanglement. Conclusions are given in section IV.

\section{Entanglement detection of three-qubit pure state}

Any pure
three-qubit states $|\psi\rangle$ can be either fully separable,
biseparable or genuine entangled. A fully separable pure three-qubit
state $|\psi\rangle$ can be written as a tensor product of three pure states,
$|\psi\rangle=|\phi_1\rangle\otimes|\phi_2\rangle\otimes|\phi_3\rangle$.
While biseparable states have three different kinds depending on the
partitions. If $|\psi\rangle$ is separable under partition of the
first qubit and the rest qubits, it has the form,
$|\psi\rangle=|\phi_1\rangle\otimes|\phi_{23}\rangle$, with
$|\phi_{23}\rangle$ an entangled state of the second and the third
qubits. We denote this kind of biseparable state as $1|23$ separable
state. Analogously, there are $2|13$ and $12|3$ separable states.
These $1|23$, $2|13$ and $12|3$ separable states are biseparable ones. If state $|\psi\rangle$ is neither fully separable nor biseparable, then it
is genuine entangled. There are two kinds of genuine entangled
states under SLOCC classification
\cite{W. Dur},
$|W\rangle=\frac{1}{\sqrt{3}}(|001\rangle+|010\rangle+|100\rangle)$
and $|GHZ\rangle=\frac{1}{\sqrt{2}}(|000\rangle+|111\rangle)$.

For three-qubit mixed state $\rho$, it is fully separable if it is a
convex combination of fully separable pure states. $\rho$ is
biseparable if it can be written as the convex combination of
biseparable pure states. Furthermore, a mixed state $\rho$ is called separable under partition $1|23$ if it is a convex combination of $1|23$ separable pure state. Analogously, there are $2|13$ and $12|3$ separable mixed states. From this respect, a general biseparable state $\rho$ is also a convex combination of $1|23$, $2|13$ and $12|3$ separable mixed states. $\rho$ is genuine entangled if it is
neither fully separable nor biseparable.

We first deal with the problem of identifying bipartite
entanglement of three-qubit pure states, by realizing entanglement
measure in terms of the expectation values of local observables.
Here we adopt concurrence as the bipartite entanglement measure
\cite{W. K. Wootters, A. Uhlmann, P. Rungta, S. Albeverio2001, D. A.
Meyer, A. R. R. Carvalho}. For a bipartite pure state
$|\psi\rangle$, its concurrence is defined by
$C(|\psi\rangle)=\sqrt{1-tr \rho_1^2}$ with
$\rho_1=tr_2(|\psi\rangle\langle\psi|)$ the reduced density matrix.
For a three-qubit state $|\psi\rangle=\sum_{i,j,k=0}^1 a_{ijk}
|ijk\rangle$, $\sum_{i,j,k=0}^1 |a_{ijk}|^2=1$, if we view it as a bipartite state
under the partition of the first qubit and the rest qubits, its
squared concurrence is given by
\begin{equation}\label{three qubit bi concurrence}
\begin{array}{rcl}
C_{1|23}^2(|\psi\rangle)&=&(\sum_{j,k=0}^1
|a_{0jk}|^2)(\sum_{j,k=0}^1 |a_{1jk}|^2)\\[3mm]
&&-|\sum_{j,k=0}^1 a_{0jk}a_{1jk}^*|^2.
\end{array}
\end{equation}
After a lengthy calculation, we get that the right hand side of Eq.
(\ref{three qubit bi concurrence}) can be expressed as the quadratic
polynomial of the expectation values of Pauli operators,
\be\label{three qubit bi concurrence-1}
\ba{l}
\langle G_1 \rangle_{|\psi\rangle\langle\psi|}\equiv\frac{1}{16}(3
-\langle I I \sigma_3\rangle^2-\langle I  \sigma_3 I\rangle^2-3 \langle \sigma_3 I I \rangle^2 \\[3mm]
~~~+\langle  \sigma_3 \sigma_3 I \rangle^2 + \langle  \sigma_3 I \sigma_3 \rangle^2
- \langle I \sigma_3 \sigma_3 \rangle^2 + \langle  \sigma_3 \sigma_3 \sigma_3 \rangle^2 \\[3mm]
~~~-3 \langle  \sigma_1 I I \rangle^2 + \langle  \sigma_1 I \sigma_3 \rangle^2
+\langle  \sigma_1 \sigma_3 I \rangle^2+\langle  \sigma_1 \sigma_3 \sigma_3 \rangle^2 \\[3mm]
~~~-3 \langle  \sigma_2 I I \rangle^2 + \langle  \sigma_2 I \sigma_3 \rangle^2
+\langle  \sigma_2 \sigma_3 I \rangle^2+\langle  \sigma_2 \sigma_3 \sigma_3 \rangle^2),
\ea
\ee
where $\sigma_1=|0\rangle\langle1|+|1\rangle\langle0|$, $\sigma_2=i(|0\rangle\langle1|-|1\rangle\langle0|)$
and $\sigma_3=|0\rangle\langle0|-|1\rangle\langle1|$ are Pauli operators, $I$ is the identity operator, $\langle I I \sigma_3\rangle$ stands for
$\langle I\otimes I \otimes\sigma_3\rangle$ and so on.

By permutation we can similarly get the squared concurrence $C_{2|13}^2(|\psi\rangle)$ of $|\psi\rangle$, denoted by
\be\label{three qubit bi concurrence-2}
\ba{l}
\langle G_2 \rangle_{|\psi\rangle\langle\psi|}=\frac{1}{16}(3
-\langle I I \sigma_3\rangle^2-\langle \sigma_3 I I\rangle^2-3 \langle I\sigma_3 I \rangle^2 \\[3mm]
~~~+\langle  \sigma_3 \sigma_3 I \rangle^2 + \langle I  \sigma_3 \sigma_3 \rangle^2
- \langle  \sigma_3 I\sigma_3 \rangle^2 + \langle  \sigma_3 \sigma_3 \sigma_3 \rangle^2 \\[3mm]
~~~-3 \langle I  \sigma_1 I \rangle^2 + \langle   I \sigma_1 \sigma_3 \rangle^2+\langle \sigma_3 \sigma_1  I \rangle^2
+\langle  \sigma_3 \sigma_1  \sigma_3 \rangle^2 \\[3mm]
~~~-3 \langle I  \sigma_2 I \rangle^2 + \langle I  \sigma_2 \sigma_3 \rangle^2+\langle \sigma_3 \sigma_2  I \rangle^2
+\langle \sigma_3 \sigma_2 \sigma_3 \rangle^2),
\ea
\ee
and the squared concurrence $C_{3|12}^2(|\psi\rangle)$ of $|\psi\rangle$,
\be\label{three qubit bi concurrence-3}
\ba{l}
\langle G_3 \rangle_{|\psi\rangle\langle\psi|}=\frac{1}{16}(3
-\langle\sigma_3 I I \rangle^2-\langle I  \sigma_3 I\rangle^2-3 \langle  I I \sigma_3 \rangle^2 \\[3mm]
~~~+\langle  I \sigma_3  \sigma_3 \rangle^2 + \langle  \sigma_3 I \sigma_3 \rangle^2
- \langle  \sigma_3 \sigma_3 I\rangle^2 + \langle  \sigma_3 \sigma_3 \sigma_3 \rangle^2 \\[3mm]
~~~-3 \langle   I I \sigma_1 \rangle^2 + \langle  \sigma_3 I \sigma_1 \rangle^2
+\langle I \sigma_3  \sigma_1\rangle^2+\langle  \sigma_3 \sigma_3 \sigma_1 \rangle^2 \\[3mm]
~~~-3 \langle   I I \sigma_2 \rangle^2 + \langle  \sigma_3 I \sigma_2 \rangle^2
+\langle  I  \sigma_3 \sigma_2 \rangle^2+\langle  \sigma_3 \sigma_3 \sigma_2 \rangle^2).
\ea
\ee

Eqs. (\ref{three qubit bi concurrence-1}), (\ref{three qubit bi
concurrence-2}) and (\ref{three qubit bi concurrence-3}) give a
realization of experimental measurement of bipartite entanglement
of three-qubit pure states. One can obtain the value of
concurrence by measuring the expectation values of Pauli operators. If $\langle G_i \rangle_{|\psi\rangle\langle\psi|}>0$, then
three-qubit pure state $|\psi\rangle$ is not separable between the $i$-th qubit and
the rest. If $\langle G_i \rangle_{|\psi\rangle\langle\psi|}=0$, then
the three-qubit pure state $|\psi\rangle$ is at least biseparable, $i=1,2,3$.

Note that any three-qubit pure state $|\psi\rangle$ is fully
separable if and only if its concurrence under all biseparable partitions
are zero. $|\psi\rangle$ is biseparable if and only if its
concurrence between one fixed qubit and the rest two qubits is zero, while the other two bipartite
concurrence are not zero. At last, $|\psi\rangle$ is genuine entangled
if and only if its concurrence for all bipartite partitions
are nonzero. Therefore, employing the nonlinear operators
$G_j$, $j=1,2,3$, we have the following result for experimentally
identifying different kinds of entanglement in arbitrary unknown three-qubit pure states.

\begin{theorem}\label{three pure detect}
For any pure three-qubit state $|\psi\rangle$, we have\\
(i) $|\psi\rangle$ is fully separable if and only if
$\langle G_j\rangle_{|\psi\rangle\langle\psi|}=0$, for $j=1,2$, or $j=2,3$, or $j=1,3$. \\
(ii) $|\psi\rangle$ is separable between the $i$-th qubit and the rest if and only if
$\langle G_i\rangle_{|\psi\rangle\langle\psi|}=0$ and
$\langle G_j\rangle_{|\psi\rangle\langle\psi|}>0$, $j\in\{1,2,3\}$ and $j\neq i$, $i=1,2,3$. \\
(iii) $|\psi\rangle$ is genuine entangled if and only if
$\langle G_j\rangle_{|\psi\rangle\langle\psi|}>0$, $j=1,2$, or $j=2,3$, or $j=1,3$.
\end{theorem}

In fact, to determine the type of entanglement existed in three-qubit pure state, one can resort to the Schmidt decomposition across the bipartition $1|23$, $2|13$ and $12|3$, and then conclude whether it is fully separable, biseparable or genuine entangled. However, this method works in theory and requires that we have already known precisely all coefficients of the pure state. In contrast, Theorem \ref{three pure detect} works for any unknown pure three-qubit states
and it is operational experimentally.

From the view of entanglement witness, $G_i$, $i=1,2,3$, can be
regarded as nonlinear entanglement witness operators.
Theorem \ref{three pure detect} shows that there exist a complete set of
entanglement witnesses to identify all kinds of possible pure three-qubit entanglement:
fully separable, three types of biseparable entanglement and genuine entangled states.
Compared with usual Bell inequality, which requires
infinitely many measurements of observables, our local operators are fixed. In another word, to detect and differentiate pure three-qubit entanglement, one only needs to measure the coincidence probabilities: $\sigma_3\otimes\sigma_3\otimes\sigma_3$, $\sigma_3\otimes\sigma_3\otimes\sigma_1$, $\sigma_3\otimes\sigma_3\otimes\sigma_2$, $\sigma_3\otimes\sigma_1\otimes\sigma_3$, $\sigma_3\otimes\sigma_2\otimes\sigma_3$, $\sigma_1\otimes\sigma_3\otimes\sigma_3$, $\sigma_2\otimes\sigma_3\otimes\sigma_3$ in $G_1$, $G_2$ and $G_3$.
These finite and deterministic measurements make the experimental entanglement detection simpler.

For
multipartite high dimensional pure state, there are many different
kinds of entanglement. For example, an $N$-partite system can be
genuine entangled, ${N\choose2}$
different biseparable, ${N\choose3}$ different tripartite
separable, $\cdots$, fully separable. Different kinds of
entanglement could be detected by expanding the
${N\choose2}$ different bipartite concurrence, ${N\choose3}$ different tripartite generalized concurrence, $\cdots$, and ${N\choose{N-1}}$ different $N-1$ partite
generalized concurrence in terms of the expectation values of Hermitian operators.
Hence all pure state entanglement could be detected completely by measuring the
expectation values of local observables.

\section{Entanglement detection of three-qubit mixed states}

To detect entanglement of mixed states is much more complicated.
Even for a known mixed state, one has no general approach to judge its separability.
In order to identify the entanglement of a three-qubit mixed state $\rho$, here we give two kinds inequalities to detect three-qubit entanglement in terms of the expectation values of complementary observables. First, let $A$, $B$, and $C$ denote the observables acting on the first, second and third qubits respectively.
$\{A_i=\vec{a}_i\cdot\vec{\sigma}\}_{i=1}^3$,
$\{B_j=\vec{b}_j\cdot\vec{\sigma}\}_{j=1}^3$ and $\{C_k=\vec{c}_k\cdot\vec{\sigma}\}_{k=1}^3$ are arbitrary complete
set of complementary observables with the same orientations,
$\vec{\sigma}$ is the vector composed by Pauli operators
\cite{S. Yu}. For a set of three mutual complementary
observables $\{A_i\}_{i=1}^3$, we denote $\mu_A=-iA_1A_2A_3$ as its
orientation which can assume only two values $\pm 1$. If $\mu_A=1$
the orientation of the basis formed by the three real vectors
$\vec{a}_i$ is right-handed, the same definition of orientation applies to
$\vec{\sigma}$. The orientations of $\{B_j\}_{j=1}^3$ and $\{C_k\}_{k=1}^3$ are defined
similarly. In the following when we refer to the complementary
observables $\{A_i\}_{i=1}^3$, $\{B_j\}_{j=1}^3$ and $\{C_k\}_{k=1}^3$ we mean that they
have the same orientations as the default.

\begin{theorem}\label{th mixed-1}
For any three-qubit mixed state $\rho$,
if it is separable under the partition $1|23$ and $12|3$, then it satisfies
\begin{equation}
\begin{array}{rcl}
\langle T_1\rangle_\rho=&&\langle 1+ B_3 + A_3 C_3 + A_3 B_3 C_3 \rangle_\rho^2\\
&&-\langle C_3 + B_3 C_3 +A_3+ A_3 B_3 \rangle_\rho^2\\
&& -\langle A_1 C_1 +A_1 B_3 C_1 +A_2 C_2 +A_2 B_3 C_2\rangle_\rho^2 \geq 0,
\end{array}
\end{equation}
for all complementary local observables.
\end{theorem}

The proof of Theorem \ref{th mixed-1} can be derived analogously in the light of the second part of the proof of the main result in Ref. \cite{mj-2byd inequ}. This Theorem tells us that if $\langle T_{1}\rangle_{\rho}< 0$
for some complementary local observables, then quantum state $\rho$ is not separable under the partition $1|23$ and $12|3$ and it is surely entangled. Similarly, if $\rho$ is separable under the partition $2|13$ and $12|3$, then it satisfies
$\langle T_2\rangle_\rho=\langle 1+ A_3 +B_3 C_3  + A_3 B_3 C_3 \rangle_\rho^2
-\langle C_3 + A_3 C_3 +B_3+ A_3 B_3 \rangle_\rho^2
-\langle B_1 C_1 +A_3 B_1 C_1 +B_2 C_2+ A_3 B_2 C_2\rangle_\rho^2 \geq0$
for all complementary local observables and if $\rho$ is separable under the partition $1|23$ and $2|13$, then it satisfies
$\langle T_3\rangle_\rho=\langle 1+ C_3 + A_3 B_3 + A_3 B_3 C_3 \rangle_\rho^2
-\langle B_3 + B_3 C_3 +A_3+ A_3 C_3 \rangle_\rho^2
-\langle A_1 B_1 +A_1 B_1 C_3 +A_2 B_2+ A_2 B_2 C_3\rangle_\rho^2 \geq0 $
for all complementary local observables.
Next, we illustrate the capability of Theorem \ref{th mixed-1} in detecting entanglement by some examples.

Example 1. For quantum state
\begin{eqnarray*}
\rho_1=&&\frac{1}{3}(|\psi^+\rangle\langle\psi^+|_{AB}\otimes |0\rangle\langle0|_C + |\psi^+\rangle\langle\psi^+|_{AC}\otimes |0\rangle\langle0|_B \\&&+ |\psi^+\rangle\langle\psi^+|_{BC}\otimes |0\rangle\langle0|_A),
\end{eqnarray*}
with $|\psi^+\rangle=\frac{1}{\sqrt{2}}(|00\rangle+|11\rangle)$, it has $\langle T_{1}\rangle_{\rho_1}=-\frac{16}{9}$ if we take $B_i=C_i=\sigma_i$, and $A_i=U_1\sigma_iU^\dagger_1$ with $U_1= |0\rangle\langle 1| -|1\rangle\langle 0|$, $i=1,2,3$. So $\rho_1$ is identified as an entangled state by Theorem \ref{th mixed-1}.

Example 2. For quantum state
\begin{eqnarray}\label{ex three qubit ppt }
\sigma_b=\frac{7b}{7b+1}\sigma_{insep} + \frac{1}{7b+1} |\phi_b\rangle
\langle \phi_b|,
\end{eqnarray}
where
$$
\begin{array}{cl}
\sigma_{insep}=
\frac{2}{7}(|\psi_1\rangle \langle \psi_1| + |\psi_2\rangle \langle
\psi_2|+ |\psi_3\rangle \langle \psi_3|)+\frac{1}{7}|011\rangle
\langle 011|,\\[2mm]
|\phi_b\rangle= |1\rangle\otimes
(\sqrt{\frac{1+b}{2}}|00\rangle + \sqrt{\frac{1-b}{2}}|10\rangle ),\\[2mm]
|\psi_1\rangle= \frac{1}{\sqrt{2}}(|000\rangle +
|101\rangle),\\[2mm]
|\psi_2\rangle= \frac{1}{\sqrt{2}}(|001\rangle +
|110\rangle),\\[2mm]
|\psi_3\rangle= \frac{1}{\sqrt{2}}(|010\rangle +
|111\rangle),
\end{array}
$$
$\sigma_b$ is entangled and positive under arbitrary partial transposition for $0\leq b\leq 1$. So it is a PPT entangled state in three-qubit system. Now if we choose $A_i=U_2\sigma_iU^\dagger_2$, $B_i=V_2\sigma_iV^\dagger_2$, $C_i=\sigma_i$, with $U_2= |0\rangle\langle 1| - |1\rangle\langle 0| $, $V_2=\frac{1}{\sqrt{2}}( |0\rangle\langle 0| + |0\rangle\langle 1| - |1\rangle\langle 0| + |1\rangle\langle 1|)$, $i=1,2,3$, then $\langle T_{1}\rangle_{\sigma_b}=-\frac{32 b (-1+b+\sqrt{1-b^2})}{(1+7 b)^2}<0$ for $0< b< 1$. Therefore, Theorem \ref{th mixed-1} has advantages in PPT entanglement detection in three-qubit system.

Example 3. For the quantum state
\begin{eqnarray*}
\rho_3=p \sigma_b +\frac{1-p}{8}I,
\end{eqnarray*}
with $0\leq p,\  b\leq 1$, which is a mixture of PPT state $\sigma_b$ in Eq. (\ref{ex three qubit ppt }) with white noise, $\rho_3$ is still a PPT state with two parameters $b$ and $p$. Below we plot the expectation value $\langle T_{1}\rangle_{\rho_3}$ with the help of local observables given in Example 2 (See Fig. 1). The dark (blue) region in the contour plot represents the PPT entangled state $\rho_3$ that Theorem \ref{th mixed-1} could detect.
\begin{center}
\begin{figure}[!h]\label{fig}
\resizebox{6cm}{!}{\includegraphics{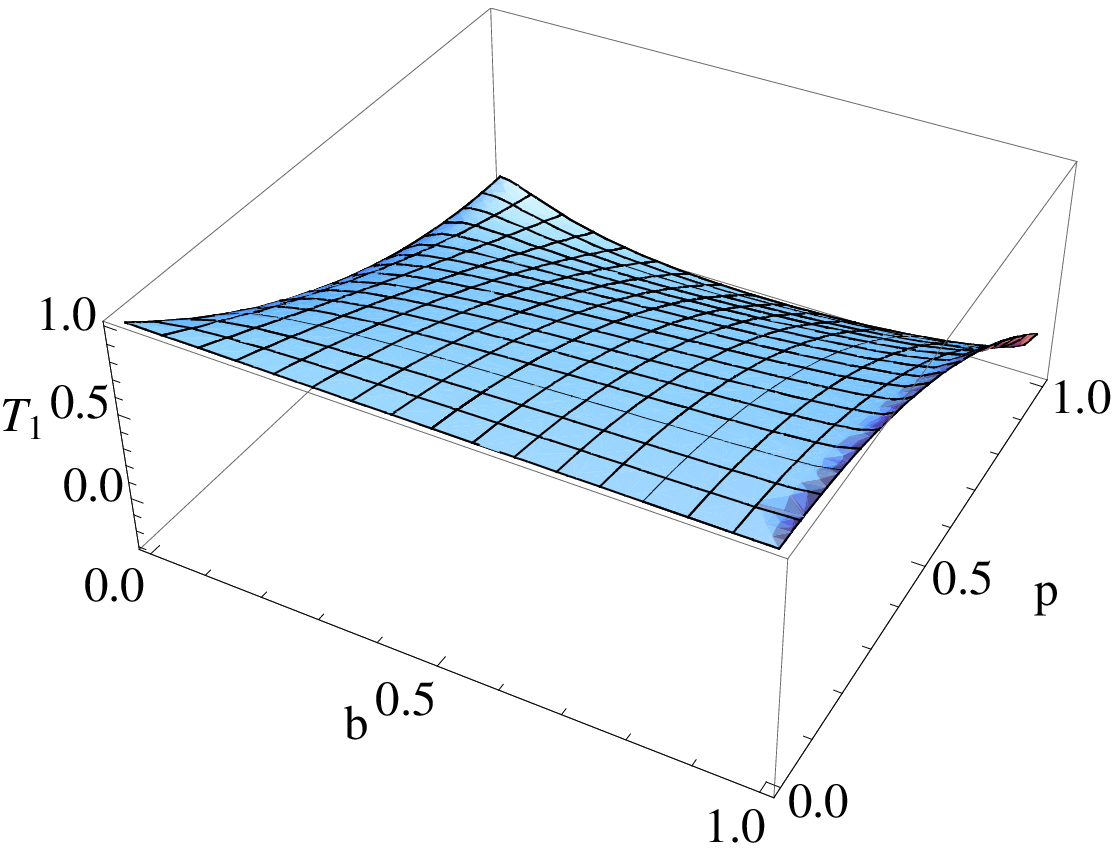}}\\[2mm]
\resizebox{5cm}{!}{\includegraphics{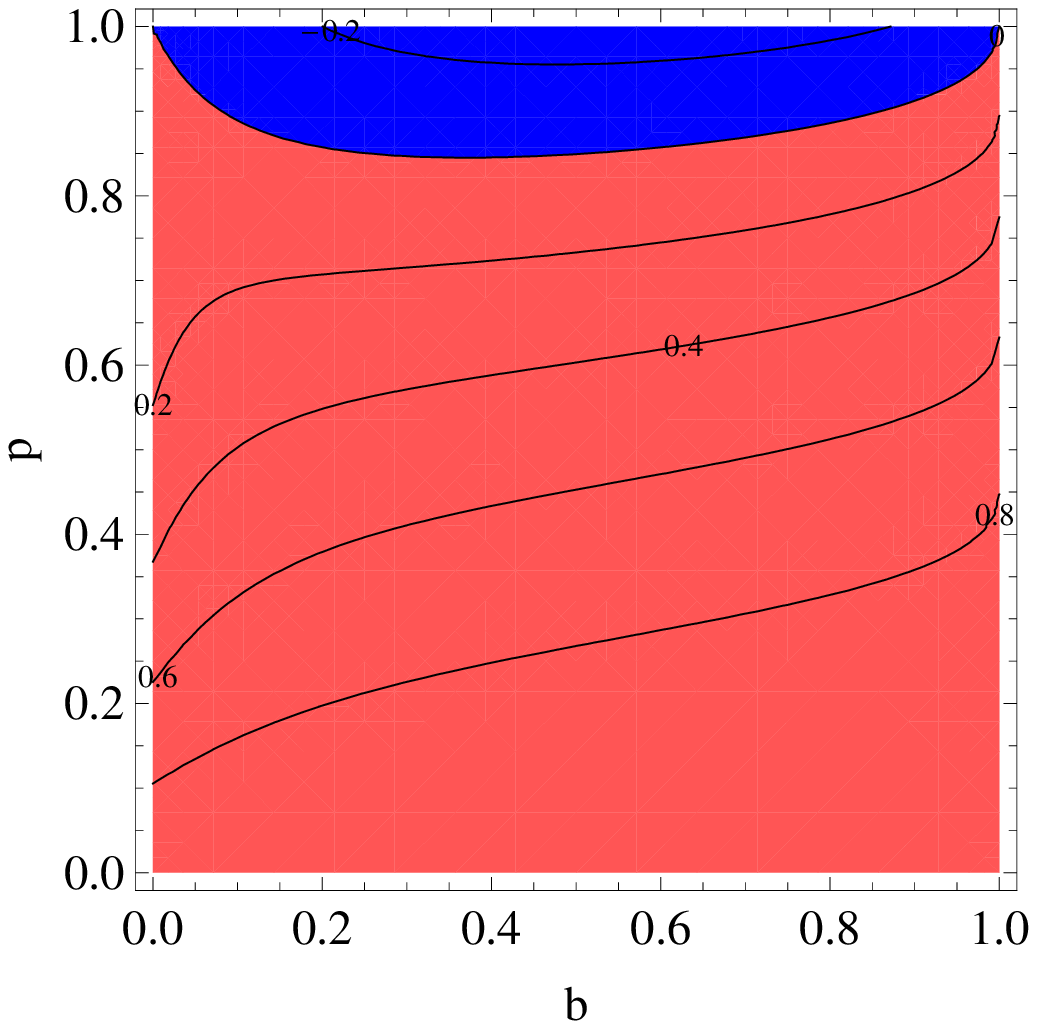}}\caption{(Color online) Here we select $A_i=U_2\sigma_iU^\dagger_2$, $B_i=V_2\sigma_iV^\dagger_2$, $C_i=\sigma_i$, with $U_2= |0\rangle\langle 1| - |1\rangle\langle 0| $, $V_2=\frac{1}{\sqrt{2}}( |0\rangle\langle 0| + |0\rangle\langle 1| - |1\rangle\langle 0| + |1\rangle\langle 1|)$, $i=1,2,3$. The first plot describes expectation value $\langle T_{1}\rangle_{\rho_3}$ with respect to $p$ and $b$. The second plot is the contour plot of the first.}
\end{figure}
\end{center}

Now we propose another kind of inequality to identify different entanglement in three-qubit system. Let
\begin{equation}\label{eq f}
\begin{array}{rcl}
\langle F_1\rangle_\rho&&=\langle 1+B_3 C_3 \rangle_\rho^2
-\langle B_3 +C_3\rangle_\rho^2-\langle B_1 C_1+ B_2 C_2\rangle_\rho^2,\\[3mm]
\langle F_2\rangle_\rho&&=\langle 1+A_3 C_3\rangle_\rho^2-\langle A_3+C_3\rangle_\rho^2-\langle A_1 C_1+A_2 C_2\rangle_\rho^2,\\[3mm]
\langle F_3\rangle_\rho&&=\langle 1+A_3B_3\rangle_\rho^2-\langle A_3+B_3\rangle_\rho^2-\langle A_1B_1+A_2B_2\rangle_\rho^2,
\end{array}
\end{equation}
then we have the following result.

\begin{theorem}\label{th mixed}
For any three-qubit mixed state $\rho$, we have\\
(i) if it is fully separable, then it satisfies $\langle F_l\rangle_{\rho}\geq0$
for all complementary local observables $\{A_i\}_{i=1}^3$, $\{B_j\}_{j=1}^3$ and $\{C_k\}_{k=1}^3$, $l=1,2,3$. \\
(ii) if it is biseparable, then
\begin{eqnarray}\label{system of inequ for bisep mixed}
\sum_{l=1}^3 \langle F_l\rangle_{\rho}\geq-2,
\end{eqnarray}
for all complementary local observables $\{A_i\}_{i=1}^3$, $\{B_j\}_{j=1}^3$ and $\{C_k\}_{k=1}^3$. \\
(iii) if it violates inequality (\ref{system of inequ for bisep mixed}), then it is genuine entangled.
\end{theorem}

Proof. Let $\rho=\sum_k p_k |\psi_k\rangle\langle\psi_k|$, $\sum_k p_k =1$,
$p_k\geq 0$ be an arbitrary three-qubit mixed state.

(i) If $\rho$ is fully separable, then $\{|\psi_k\rangle\}$ are all fully separable,
at least in one such pure state decomposition. Hence the reduced bipartite state
$\rho_{\overline{l}}^{(k)}=Tr_{l}(|\psi_k\rangle\langle\psi_k|)$ is also
separable for all $k$, $l=1,2,3$, and $\overline{l}$ denotes the absence of $l$
in the set $\{1,2,3\}$. In the light of the main results in
Ref. \cite{S. Yu}, one gets $\langle
F_l\rangle_{|\psi_k\rangle\langle\psi_k|}\geq0$ if and only if
$\rho_{\overline{l}}$ is separable. Therefore, if
$|\psi_k\rangle$ is fully separable, then $\langle
F_l\rangle_{|\psi_k\rangle\langle\psi_k|}\geq0$, $l=1,2,3$,
$\forall k$. Note that if $a_i^2 \geq b_i^2 +c_i^2 +x$ holds for
arbitrary real numbers $b_i$ and $c_i$, nonnegative $a_i$ and $x$,
$i=1, \cdots, n$, then $(\sum_{i=1}^n p_i a_i)^2 \geq (\sum_{i=1}^n
p_i b_i)^2 +(\sum_{i=1}^n p_i c_i)^2 +x$ for $0 \leq p_i \leq 1$ and
$\sum_{i=1}^n p_i=1$. This observation makes $\langle
F_l\rangle_{\rho}\geq0$, $l=1,2,3$, for all complementary local observables and
for fully separable quantum state $\rho$.

(ii) Suppose $\{|\psi_k\rangle\}$ are all biseparable. Without loss of
generality, we assume $|\psi_1\rangle$ is $1|23$ separable, then it
satisfies $\langle
F_2\rangle_{|\psi_1\rangle\langle\psi_1|}\geq0$ and $\langle
F_3\rangle_{|\psi_1\rangle\langle\psi_1|}\geq0$ for all
complementary local observables. Taking into account that the
minimum of $\langle F_1\rangle_{|\psi_1\rangle\langle\psi_1|}$
for arbitrary complementary local observables is two times minimal
eigenvalue of the partial transposed matrix of $\rho^{(1)}_{23}$
\cite{S. Yu}, one has $\langle
F_1\rangle_{|\psi_1\rangle\langle\psi_1|}\geq-2$ for all
complementary local observables. Therefore we get $\sum_{l=1}^3 \langle
F_l\rangle_{|\psi_k\rangle\langle\psi_k|}\geq-2$ for all
complementary local observables and for arbitrary biseparable state
$|\psi_k\rangle$.
Consequently, for biseparable mixed state $\rho$,
we have $\sum_{l=1}^3 \langle F_l\rangle_{\rho}\geq-2$ for all
complementary local observables.

(iii) The result is obvious.\qed

As an example, we consider a state of mixture of $|W\rangle =\frac{1}{\sqrt{3}}(|001\rangle+|010\rangle+|100\rangle)$ with white
noise, $\rho_w=p|W\rangle\langle W|+\frac{1-p}{8}I$, $0\leq p \leq 1$. It is
detected as genuine entangled by the entanglement witness in
Ref. \cite{M. Bourennane} when $p>0.62$. If we take
$A_k=B_k=C_k=\sigma_k$ in the local operators
in Eq. (\ref{eq f}), $k=1,2,3$, we have $\sum_{l=1}^3 \langle
F_l\rangle_{\rho_w}>-2$ when $p>0.92$ and $\langle
F_l\rangle_{\rho_w}<0$ when $p>0.56$, $l=1,2,3$. Hence we know
that $\rho_w$ is genuine entangled when $p>0.92$ and entangled when
$p>0.56$ by Theorem \ref{th mixed}. Our entanglement witness
can detect entanglement of $\rho_w$ better. Although we
can not detect all genuine entanglement in the mixed state
$\rho_{w}$ by Theorem \ref{th mixed}, the advantage of our method
here is that it may be experimentally implemented.

\section{Conclusions}

As a summary, we have expressed the bipartite
entanglement of three-qubit pure states in the form of expectation
values of Pauli operators. With the aid of this expression, we
have solved the problem of entanglement identification for three-qubit pure states
completely, by giving the necessary and sufficient conditions for
fully separable states, biseparable states and genuine entangled
states. This approach can be
generalized to multipartite high dimensional cases. Therefore one
can recognize the separability of pure states both theoretically and
experimentally. Additionally, we have also derived two inequalities in the form of the expectation values of complementary observables to detect PPT entanglement and genuine entanglement for three-qubit mixed state respectively. This results may help experimental
entanglement detection and identification.

\end{document}